\documentclass[aps,pra,amsmath,twocolumn,groupedaddress,superscriptaddress]{revtex4-1}
 \usepackage{graphicx}
 \usepackage{color}
 \usepackage{graphics}
 \usepackage{nicefrac}
 \usepackage{amsfonts}
 \usepackage{ulem}

 \usepackage{amssymb}
 \usepackage{amsmath}

 \usepackage{subfigure}
 \usepackage{multirow}
 \usepackage{tabularx}
 \usepackage{array}

 \usepackage{units}

 \usepackage{tensor}
 \usepackage{braket}

 \usepackage{bm}
 \usepackage{hyperref}
\hypersetup{colorlinks=true, linkcolor=blue, citecolor=blue, urlcolor=blue}

 \providecommand{\U}[1]{\protect\rule{.1in}{.1in}}
 
\begin{document}
\title{Thermal gradient driven domain wall dynamics}
\author{M. T. Islam}
\affiliation{Physics Department, The Hong Kong University of
Science and Technology, Clear Water Bay, Kowloon, Hong Kong}
\affiliation{Physics Discipline, Khulna University, Khulna, Bangladesh}
\author{X. S. Wang}
\affiliation{School of Electronic Science and Engineering
and State Key Laboratory of Electronic Thin Film and
Integrated Devices, University of Electronic Science and Technology
of China, Chengdu 610054, China}
\affiliation{Center for Quantum Spintronics, Department of Physics,
Norwegian University of Science and Technology, NO-7491 Trondheim, Norway}
\author{X. R. Wang }
\email{[Corresponding author:]phxwan@ust.hk}
\affiliation{Physics Department, The Hong Kong University of
Science and Technology, Clear Water Bay, Kowloon, Hong Kong}
\affiliation{HKUST Shenzhen Research Institute, Shenzhen 518057, China}
\begin{abstract}
The issue of whether a thermal gradient acts like a magnetic field or
an electric current in the domain wall (DW) dynamics is investigated.
Broadly speaking, magnetization control knobs can be classified as energy-driving
or angular-momentum driving forces. DW propagation driven by a static magnetic field
is the best known example of the former in which the DW speed is proportional to the
energy dissipation rate, and the current-driven DW motion is an example of the latter.
Here we show that DW propagation speed driven by a thermal gradient can be fully
explained as the angular momentum transfer between thermally generated spin current
and DW. We found DW-plane rotation speed increases as DW width decreases.
Both DW propagation speed along the wire and DW-plane rotation speed
around the wire decrease with the Gilbert damping. These facts are consistent
with the angular momentum transfer mechanism, but are distinct from the
energy dissipation mechanism.
We further show that magnonic spin-transfer torque (STT) generated by
a thermal gradient has both damping-like and field-like components.
By analyzing DW propagation speed and DW-plane rotational speed, the coefficient
(\( \beta\)) of the field-like STT arising from the non-adiabatic process, is obtained.
It is found that  \( \beta\) does not depend on the thermal gradient; increases with
uniaxial anisotropy \(K_{\parallel}\) (thinner DW); and decreases with the damping,
in agreement with the physical picture that a larger damping or a thicker DW
leads to a better alignment between the spin-current polarization and the local
magnetization, or a better adiabaticity.	
\end{abstract}

\maketitle
\section{INTRODUCTION}
Manipulating domain walls (DW) in magnetic nanostructures has attracted much
attention because of its potential applications in data storage technology
\cite{parkin} and logic gates \cite{DA2005}. The traditional DW control knobs, namely
magnetic fields and spin-polarized currents, have certain drawbacks in applications.
In the magnetic-field-driven DW motion, energy dissipation is the main cause of DW
propagation whose speed is proportional to the energy dissipation rate
\cite{xrw1,Fe2009}, and the magnetic field tends to destroy unfavorable domains and
DWs, instead of driving a series of DWs synchronously \cite{DA2003,GS2005,MH2006}.
An electrical current drives a DW to move mainly through the angular momentum transfer
so that it pushes multiple DWs \cite{LB1996,Slonczewski,SZ2004,GT2004} in the same
direction. To achieve a useful DW speed, it requires high electrical current
densities that result in a Joule heating problem \cite{Jole2,Expecom1,Expecom2}.
To avoid these problems, spin-wave spin current has been proposed as
a more energy-efficient control parameter \cite{XSW11,NOWAK11,KOVALEV12,MagSTT}.
Thermal gradient, a way to generate spin-wave spin current, is an alternative
control knob of the DW motion. The investigation on thermal-gradient-driven
domain wall motion is meaningful not only for conventional applications, but
also for the understanding of spin wave and domain wall dynamics
\cite{NOWAK11,KOVALEV12,Jiang13,XSW14,NOWAK14,XGWANG16}, as well as for possible
recycling of waste heat \cite{spincal, Nat.com2017}.

To understand the mechanism behind thermal-gradient-driven DW dynamics,
there are microscopic theories \cite{XSW11,NOWAK11,KOVALEV12,XGWANG12,PYAN2015}
and macroscopic thermodynamic theories \cite{XSW14,NOWAK14}.
Briefly speaking, the microscopic theories suggest that magnons populated in the
hotter region diffuses to the colder region to form a magnon spin current.
The magnon spin current passes through a DW and exerts a torque on the
DW by transferring spin angular momentum to the DW. Thus, magnons drive the DW
propagating toward the hotter region of the nanowire, opposite to the magnon
current direction \cite{MagSTT,XSW11,NOWAK11}.
The thermodynamic theories anticipate that a thermal gradient generates an
entropy force which always drives the DW towards the hotter region in order to
minimize the system free energy. The macroscopic theories do not provide any
microscopic picture about DW dynamics although a thermal gradient is often
considered as an effective magnetic field to estimate DW speed \cite{XSW14,NOWAK14}
from field-driven DW theories. Thus, one interesting issue is whether a thermal
gradient in DW dynamics acts like a magnetic field or an electric current.
DW propagation speed should be sensitive to both DW width and types of a DW (transverse
DW) under an energy-driving force while the speed should be insensitive to the DW and DW
structure in the angular-momentum-driving force. This is the focus of the current work.

In this paper, we investigate DW motion along a uniaxial wire with the easy axis along
the wire direction under a thermal gradient. We found that the DW always propagates to the
hotter region with an accompanied DW-plane rotation. DW propagation speed and DW-plane
rotation speed increases as the magnetic easy-axis anisotropy and damping decreases.
We show that DW motion can be attributed to the angular momentum transfer between
magnonic spin current and the DW. Thus, we conclude that a thermal gradient interacts
with DW through angular-momentum transfer rather than through energy dissipation.
Similar to an electric current \cite{Nonad}, a thermal gradient can generate both
damping-like (or adiabatic) STT and field-like (or non-adiabatic) STT.
From the damping-dependence and anisotropy-dependence of the average DW velocity and
DW-plane rotation angular velocity, we extract  field-like STT coefficient (\(\beta\)).
It is found that \(\beta\) is independent of thermal gradient; is bigger for a thinner
DW; and decreases with the damping coefficient. We also show that in the presence of
a weak hard-axis anisotropy perpendicular to the wire, the DW still undergoes a rotating
motion. The DW propagation speed increases slightly while the DW-plane rotation speed
decreases with the strength of the hard-axis anisotropy.

\begin{figure}[!t]
 	\begin{center}
 		\includegraphics[width=8.5cm]{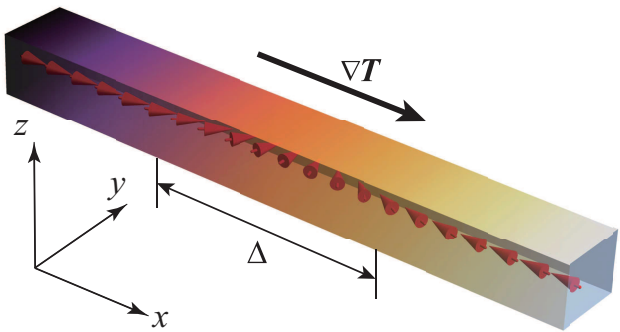}
 	\end{center}
\caption{\label{Fig1} Schematic diagram of a uniaxial magnetic nanowire
with a head-to-head DW at the center under a thermal gradient $\nabla T$.
Black (white) color represents colder (hotter) end of the sample.}
\end{figure}

\section{Model and method}
We consider a uniaxial nanowire of length $L_x$ and cross-section $L_y \times L_z$ along
the $x$-axis (easy axis) with a head-to-head DW at the center, as shown in Fig. \ref{Fig1}.
$L_y$, $L_z$ is much smaller than the DW width $\Delta$, and $\Delta$ is much smaller
than $L_x$. A thermal gradient is applied along the wire. The highest temperature is far
below the Curie temperature $T_{c}$. The magnetization dynamics is governed by
the stochastic Landau-Lifshitz-Gilbert (LLG) equation \cite{Brown1963,TLGIL},
\begin{equation}
\dfrac{d\mathbf{m}}{dt}=-\gamma \mathbf{m}\times(\mathbf{H}_{\text{eff}}+
\mathbf{h}_{\text{th}})+\alpha\mathbf{m}\times\dfrac{\partial\mathbf{m}}{\partial{t}},
 \label{sllg}
\end{equation}
where $\mathbf{m}={\mathbf{M}}/{M_{s}}$ and \(M_{s}\) are respectively the magnetization
direction and the saturation magnetization. $\alpha$ is the Gilbert damping constant and
$\gamma$ is the gyromagnetic ratio. $\mathbf{H}_\text{eff}=\frac{2A}{\mu_0M_s}\sum_\sigma
\dfrac{\partial^2\mathbf{m}}{\partial x^2_\sigma}+\frac{2K_\parallel}{\mu_0M_s} m_x
\hat{\mathbf{x}}+\mathbf{h}_\text{dipole}$ is the effective field, where $A$ is the exchange
constant, $x_\sigma$ ($\sigma=1,2,3$) denote Cartesian coordinates $x$, $y$, $z$, $K_\parallel$
is the easy-axis anisotropy, and $\mathbf{h}_\text{dipole}$ is the dipolar field.
\(\mathbf{h}_{\text{th}}\) is the stochastic thermal field.

The stochastic LLG equation is solved numerically by MUMAX3 package \cite{Mumax} in which
we use adaptive Heun solver. To balance stability and efficiency, we choose the time step
\(10^{-14}\) s with the cell size (\(2\times2\times2\)) \( \text{nm}^{3}\).
Magnetic charges at the two ends of the wire are removed to avoid their attraction to the DW.
The saturation magnetization
$M_{\text{s}} = 8 \times10^{5} \text{A}/\text{m}$ and exchange constant
$A = 13\times10^{-12}$ \(\text{J}/\text{m}\) are used to mimic permalloy in our simulations.
The thermal field follows the Gaussian process characterized by following statistics \cite{Nowak2008}
\begin{equation}
 \begin{gathered}
 \langle h_{\text{th},ip}(t)\rangle = 0,\\
\langle h_{\text{th},ip}(t)h_{\text{th},jq}(t+\Delta t)\rangle = \dfrac{2k_{\text{B}}T_{i}\alpha_{i}}
{\gamma \mu_0 M_{\text{s}} a^{3}}\delta_{ij}\delta_{pq}\delta (\Delta t),
 \end{gathered}
 \end{equation}
where \(i\) and \(j\) denote the micromagnetic cells, and \(p\), \(q\) represent the Cartesian
components of the thermal field. \(T_{i}\) and \(\alpha_{i}\) are respectively temperature
and the Gilbert damping at cell \(i\), and $a$ is the cell size. \(k_{\text{B}}\) is the
Boltzmann constant \cite{Brown1963}. The numerical results presented in this study are averaged
over 15 random configurations (for DW velocity) and 4000-5000 random configurations (for spin current).

Under the thermal gradient $\nabla_{x}T$, magnetization at different positions deviate
from their equilibrium directions differently and small transverse components \(m_{y}\)
and \(m_{z}\) are generated. The transverse components vary spatial-temporally and depend
on the local temperature. This variation generates a magnonic spin current \cite{NOWAK11}.
This magnonic spin current can interact with spin textures such as DWs. In the absence
of damping (the thermal field also vanishes), the spin current along the \(x\) direction
can be defined from the spin continuity equation derived from Eq. \eqref{sllg} as follows
\cite{XSW11},
 \begin{equation}
 \begin{split}
\dfrac{\partial \mathbf{m}}{\partial t} &=-\dfrac{1}{1+\alpha^{2}}\mathbf{m}\times\hat{\mathbf{x}}
 m_{x}K_\parallel-\dfrac{\partial \mathbf{J}}{\partial x},
 \end{split}
 \label{dspincurrent}
 \end{equation}
where
 \begin{equation}
  \centering
J(x)=\frac{2\gamma A}{\mu_0M_s}\mathbf{m}\times\frac{\partial \mathbf{m}}{\partial x},
 \label{spincurrent}
 \end{equation}
is the spin current density along \(x\)-direction due to the exchange interaction.
$\mathbf{J}(x)$ can be numerically calculated \cite{XSW11,XGWANG16}.
In the presence of damping as well as the thermal field, the contribution
of the damping term and the thermal term is proportional to $\alpha$, which is
relatively small. More importantly, according to the fluctuation-dissipation
theorem \cite{Brown1963}, the damping term and the thermal term should cancel
each other after average over a long time. Since the time scale of DW dynamics
is much longer than the thermal fluctuation, the combined contribution of
damping and thermal terms should be very small.

Integrating the $x-$component of Eq. \eqref{dspincurrent} over a space enclosed
the DW in the center and noticing the absence of the first term on the right,
we have
\begin{multline}
v_\text{current}=\frac{1}{2}\int_{-L_x/2}^{L_x/2}\frac{\partial m_x}{\partial t} dx
\\=-\dfrac{2 \gamma A}{\mu_{0}M_s}\big[\dfrac{1}{2}
( J_x|_\text{left}-J_x|_\text{right})].
  \label{DWVE}
  \end{multline}
where we have assumed the fluctuations in the domains are small
and the DW is not far from a symmetric one.
$J_x|_\text{left}$, $J_x|_\text{right}$
mean the $x$-components of the total spin current on the left and right sides of the DW.
The equation clearly shows that the DW propagates opposite to the spin current.
This is the theoretical DW velocity under the assumption of angular momentum
conservation, and it will be compared with the directly simulated DW velocity below.
\begin{figure}[t!]
	\centering
	\includegraphics[width=8.5 cm]{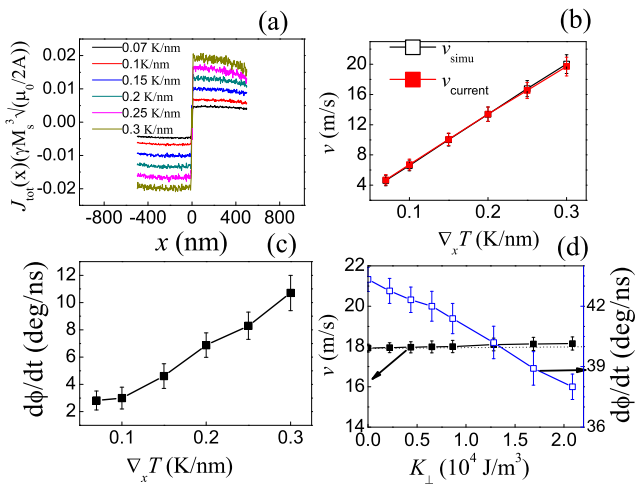}
\caption{(a) The spatial dependence of
spin current densities $J_{\text{tot}}(x)$ for various $\nabla_{x}T$.
The DW center is chosen as $x=0$. (b) Thermal gradient dependence of DW
velocity \(v_\text{simu}\) from micromagnetic simulations (open squares)
and \(v_\text{current}\) computed from total spin current (solid squares).
(c) Thermal gradient dependence of DW-plane rotation angular velocity (squares).
 In (a)(b)(c) model parameters are \(L_{x}=2048\) nm, \(L_y=L_z=4\)\ nm, \(\alpha=0.004\)
and \(K_{\parallel}=5\times10^5\quad\text{J}/\text{m}^{3}\).
(d) \(v_\text{simu}\)  (solid squares) and \(d\phi/dt\) (open squares) as a function
of $K_{\perp}$ for \(L_{x}=1024\) nm and $\nabla_x T=0.5$ K/nm.}
	\label{Fig5.1}
\end{figure}

 \section{Results}
\subsection{Average spin current and DW velocity}
To substantiate our assertion that DW propagation under a thermal gradient is through
angular-momentum effect instead of energy effect, we would like to compare the DW velocity
obtained from micromagnetic simulations and that obtained from total spin current based on
Eq. \eqref{DWVE}. Eq. \eqref{spincurrent} is used to calculate \( J_x(x)\).
Fig. \ref{Fig5.1}(a) is spatial distribution of the ensemble averaged \( J_x(x)\)
with DW  at $x=0$ for various thermal gradients. The sudden sign change of $J_x(x)$
at the DW center is a clear evidence of strong angular-momentum transfer from spin
current to the DW. Technically, magnetization of the two domains separated by the DW
point to the opposite directions, thus the spin current polarization changes its sign.
In calculating DW velocity \(v_{\text{current}}\) from Eq. \eqref{DWVE}, the spin currents
before entering DW and after passing DW are the averages of $J_x(x)$ over
$x\in [-2\Delta, -\Delta]$ and $x\in [\Delta, 2\Delta]$, where $\Delta$ is the DW width
which is 16 nm in the current case. The thermal gradient dependence of \(v_{\text{current}}\)
is shown in Fig. \ref{Fig5.1}(b) (solid squares). \(v_{\text{current}}\) compares well with
the velocity \(v_{\text{simu}}\) (open squares) obtained directly from simulations by
extracting the speed of the DW center along $x$-direction. The DW velocity is
linearly proportional to the temperature gradient $v=C\nabla_x T$, with the thermal mobility
$C=6.66\times10^{-8}$ $\text{m}^2\text{s}^{-1}\text{K}^{-1}$
for $v_\text{simu}$ or $C=6.59\times10^{-8}$ $\text{m}^2\text{s}^{-1}\text{K}^{-1}$
for $v_\text{current}$.
\begin{figure}[!t]
	\begin{center}
		\includegraphics[width=8 cm]{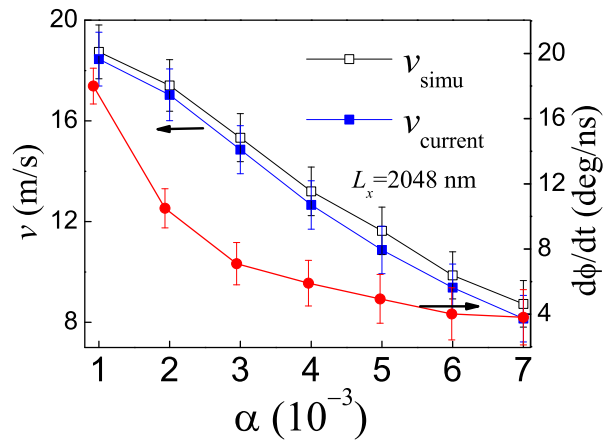}
	\end{center}
	\caption{\label{dam} Damping \(\alpha\) dependence of the DW dynamics: $v_{\text{simu}}$
		(Open squares); $v_{\text{current}}$ (solid squares ); and $d\phi/dt$ (solid circles).
		Model parameters are $\nabla_{x} T=0.2$ K/nm,
		$K_{\parallel}=5\times10^5\quad  \text{J}/\text{m}^{3}$, $L_x=2048$ nm and $L_y=L_z =4$ nm.}
\end{figure}
It is noted that \(v_{\text{current}}\) almost coincides with \(v_{\text{simu}}\) except a
small discrepancy at very high thermal gradient when the nonlinear effects is strong.
The small discrepancy may be attributed to the large fluctuations as well as the
contribution from the damping, the dipolar and stochastic fields.
These observations are consistent with magnonic STT \cite{XSW11,NOWAK11,XGWANG12,PYAN2015}.
It is observed that the DW-plane rotates around the $x$-axis counter-clockwise for head-to-head
DW and clockwise for tail-to-tail DW during DW propagation. DW rotation speed \(d\phi/dt\)
(squares) is shown in Fig. \ref{Fig5.1} (c)) as a function of $\nabla_{x}T$.

 \subsection{Damping and anisotropy dependence of DW dynamics}

An energy-effect and angular-momentum-effect have different damping-dependence and
anisotropy-dependence of DW dynamics. To distinguish the roles of energy and the
angular-momentum in thermal-gradient driven DW dynamics, it would be useful to probe how the
DW dynamics depends on $\alpha$ and \(K_{\parallel}\). Damping have two effects on the spin currents:
one is the decay of spin current during its propagation so that the amount of spin angular
momentum deposited on a DW should decrease with the increase of the damping coefficient.
As a result, the DW propagation speed and DW-plane rotation speed should also be smaller for a
larger $\alpha$. Indeed, this is what we observed in our simulations as shown in Fig. \ref{dam}(a)
for DW speed and DW-plane rotation speed (open squares for $v_{\text{simu}}$, solid circles for
$v_{\text{current}}$, and stars for $d\phi/dt$). The model parameters are \(L_x=2048\), $L_y=
L_z =4$ nm, \(\nabla_{x} T=0.2\) K/nm and  $K_{\parallel}=5\times10^5$ $\text{J}/\text{m}^{3}$.
The second damping effect is that the larger $\alpha$ helps the spin current polarization to
align with the local spin. This second effect enhances the adiabatic process that is
important for non-adiabatic STT or field-like torque discussed in the next subsection.
Therefore, $\alpha-$dependence of DW dynamics supports the origin of thermal driven
DW dynamics to be the angular-momentum effect, not the energy effect that would lead
to a larger $v_{\text{simu}}$ and $d\phi/dt$ for a larger \(\alpha\)
\cite{xrw1,Fe2009,RW2010,Euro,AM2007} instead of a decrease observed here.
 \begin{figure}[!t]
 	\begin{center}
 		\includegraphics[width=8.5 cm]{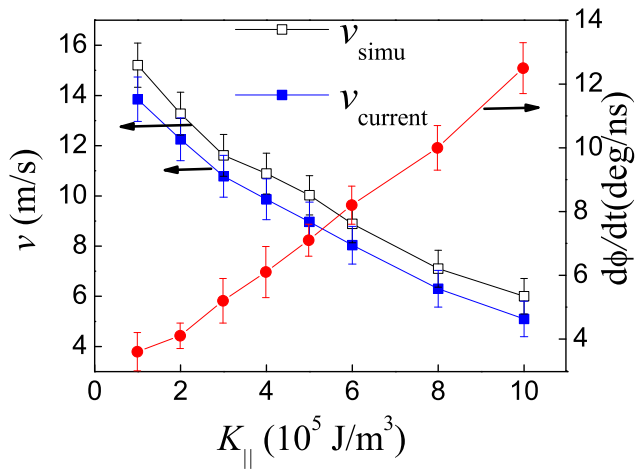}
 	\end{center}
\caption{Anisotropy $K_{\parallel}$ dependence of the DW dynamics: $v_{\text{simu}}$ (open squares);
$v_{\text{current}}$(solid squares); and $d\phi/dt$ (solid circles). Model parameters are
$L_x$= 2048 nm, $L_y$=$L_z$=4 nm, \(\alpha=0.004\) and $\nabla_{x} T=0.2$ K/nm.}
 	\label{Kude}
 \end{figure}

Here we would like to see how the DW dynamics depends on uniaxial anisotropy \(K_{\parallel}\).
Fig. \ref{Kude} shows both $v_{\text{simu}}$ (open squares), $v_{\text{current}}$ (filled
squares) and $d\phi/dt$ (circles) for $L_x$= 2048 nm,  \(\alpha=0.004\) and $\nabla_{x} T=0.2$.
The DW propagation speed, $v_{\text{simu}}$ decreases with \(K_{\parallel}\) while DW-plane
rotational speed increases with \(K_{\parallel}\). These results seem follow partially the behavior
of magnetic-field induced DW motion, in which DW propagation speed is proportional to DW width
(\(\Delta\sim\sqrt{\dfrac{A}{K_{\parallel}}}\)) or decrease with \(K_{\parallel}\), and partially electric
current driven DW motion, in which DW-plane rotational speed increases with \(K_{\parallel}\).
Thus, one may tend to conclude that a thermal gradient behaves more like a magnetic field
rather than an electric current from the DW width dependence of DW propagation speed,
opposite to our claim of the angular-momentum effects of the thermal gradient.
It turns out, this is not true. The reason is that magnon spectrum,
$\omega_{\mathbf{k}}=\frac{2\gamma}{\mu_0M_s}\left(Ak^{2}+K_{\parallel}\right)$, has a gap in a system with
magnetic anisotropy. The larger \(K_{\parallel}\) is, the bigger the energy
gap will be. Thus, it becomes harder to thermally excite magnon.
As a result, the spin current decreases as \(K_{\parallel}\) increases.
To see whether the thermal-gradient driven DW motion is due to the
angular-momentum transfer or not, one should compare whether
$v_{\text{simu}}$ and $v_{\text{current}}$ maintain a good agreement
with each other as \(K_{\parallel}\) varies. Indeed, a good agreement between
$v_{\text{simu}}$ and $v_{\text{current}}$ is shown in Fig. \ref{Kude}.
This conclusion is also consistent with existing magnonic STT theories
\cite{RW2010,Euro,AM2007}.

\subsection{Separation of adiabatic and non-adiabatic torques}
 \begin{figure}[!b]
 	\centering
\includegraphics[width=8.5cm]{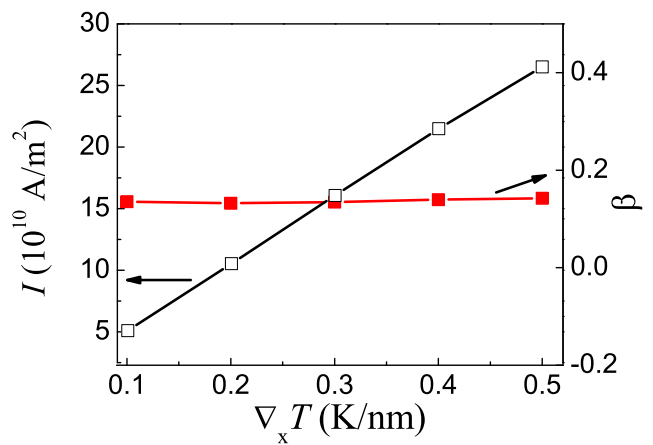}
\caption{Model parameters are $K_{\parallel}=5\times10^5$ $ \text{J}/\text{m}^{3}$, \(\alpha=0.004\),
$L_x$= 1024 and $L_y$=$L_z$=4 nm.  Effective electric current density $I$ (open squares)
and $\beta$ (solid squares) are plotted as functions of $\nabla_{x} T$.}
 	\label{beta1}
 \end{figure}

We have already demonstrated that a thermal gradient interacts with DW through magnonic STT
rather than through energy dissipation. It is then interesting to know what kind of STTs a
thermal gradient can generate. Specifically, whether a magnonic spin current generates
damping-like (adiabatic), or field-like (Non-adiabatic) torques, or both just like an
electric current \cite{Nonad} does. To extract the STT generated from a thermal gradient, we
approximate DW dynamics by the motion of its collective modes of DW center $X$ and the titled
angle $\phi$ of DW-plane. Subject to both damping-like and field-like torques, using the
travelling-wave ansatz \cite{RW2010,Euro,AM2007}, $\tan(\theta/2)=\exp[(x-X)/\Delta]$ where
\(\Delta\sim\sqrt{{A}/{K_{\parallel}}}\),  one can derive the equations for X and $\phi$,
\begin{equation}
\dfrac{\alpha}{\Delta}\dfrac{d X}{d t}+\dfrac{d\phi}{d t}=\dfrac{\beta}{\alpha}u, \quad
\dfrac{1}{\Delta}\dfrac{d X}{d t}-\dfrac{\alpha d \phi}{d t}=\dfrac{u}{\alpha}.
\end{equation}
From the above two equations, one can straightforwardly find DW propagating speed and
DW-plane rotation speed,
\begin{equation}
 v=\dfrac{(1+\alpha\beta)}{(1+\alpha^2)}u, \quad \dot\phi=\dfrac{(\beta-\alpha)}{(1+\alpha^2)}u.
 \label{velo}
\end{equation}

\begin{figure}[!t]
 	\begin{center}
 		\includegraphics[width=8.5cm]{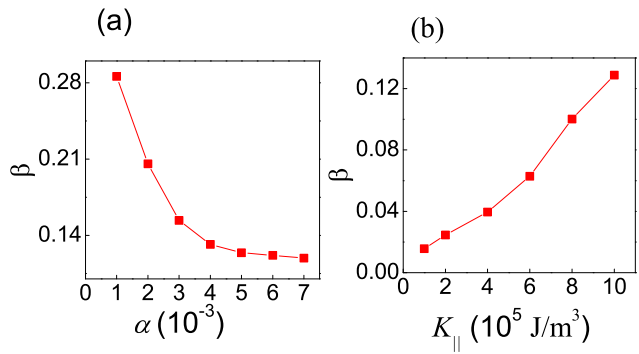}
 	\end{center}
\caption{\label{vv} Model parameters are $\nabla_{x} T$=0.5 K/nm, $L_x$= 1024 nm and $L_y$=$L_z$=4 nm.
(a) $\alpha$-dependence of $\beta$ for $K_{\parallel}=10^6$ and $ \text{J}/\text{m}^{3}$.
(b) $K_{\parallel}$-dependence of $\beta$  for $\alpha=0.004$.}
 	\label{beta2}
 \end{figure}

One can extract $\beta$ and equivalent electric current density \(I=(2eM_{s}u)/g\mu_{B}P\) from $v$
and $d\phi/dt$ obtained in simulations. For \(\alpha=0.004\), $K_{\parallel}=10^{6}$ $\text{J}/\text{m}^{3}$,
the \(I\) and $\beta$ are obtained and plotted in Fig. \ref{beta1} as a function of $\nabla_{x} T$.
It is evident that \(I\) linearly increases with $\nabla_{x} T$ and \(\beta\) is independent of
$\nabla_{x}T$ as it should be. We then fixed $\nabla_{x} T=0.5$ K/nm, and repeat simulations and
analysis mentioned above for various \(\alpha\) and $K_{\parallel}$. Fig. \ref{beta2} (a) and (b) shows
\(\beta\) as a function of \(\alpha\) and $K_{\parallel}$. From the figure, it is evident that  \(\beta\)
decreases with \(\alpha\). This is because the larger damping favors the alignment of spin current
polarization with the local spin so that the non-adiabatic effect, \(\beta\), becomes smaller.
\(\beta\) increases with \(K_{\parallel}\) for the similar reason: Larger \(K_{\parallel}\) means a thinner DW
so that it is much harder for the spin current polarization to reverse its direction after passing
through the thinner DW, i.e. a stronger non-adiabatic effect.

\begin{figure}[!t]
 \begin{center}
 	\includegraphics[width=8.5cm]{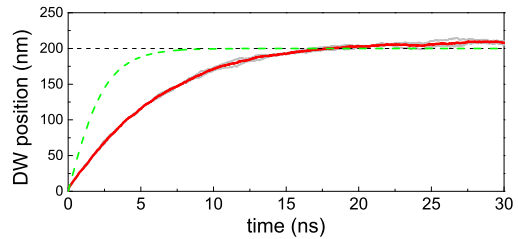}
 	\end{center}
\caption{Domain wall position versus time in a Gaussian temperature profile.
The gray lines are raw data for different random seeds and the red line
is the averaged result.
The green dashed line is theoretical result using the thermal mobility
$C=6.66\times10^{-8}$ $\text{m}^2\text{s}^{-1}\text{K}^{-1}$ obtained
from Fig. 2(b).}
 	\label{Gaussian}
 \end{figure}

In some experiments, the temperature gradient is generated by a laser spot\cite{exp2015}.
The laser spot will induce a Gaussian distribution of the temperature over the space
\cite{exp2015,Gaussian2017}. In Fig. \ref{Gaussian}, we show the DW motion in a Gaussian
temperature profile $T(x)=T_0\exp\left(-\frac{(x-x_L)^2}{2\sigma^2}\right)$ by plotting
the DW position against the time. Here we use the same parameters as those in Fig. 2(b),
except a longer wire $L_x=2048$ nm, and $T_0=400$ K, $\sigma=200$ nm, $x_L=200$ nm.
Theoretically, if the instantaneous DW speed under a Gaussian temperature is the same as
that in the constant thermal-gradient case, we should expect $\frac{dx}{dt}=C\frac{dT}{dx}$,
where the thermal mobility $C$ is the same as that in Fig. 2(b).
Using $C=6.66\times10^{-8}$ $\text{m}^2\text{s}^{-1}\text{K}^{-1}$, the above differential
equation for $x(t)$ can be numerically solved with initial condition $x(0)=0$.
The result is plotted in Fig. \ref{Gaussian} in green dashed line.
The simulated speed is smaller than this theoretical result. This is probably because,
for the constant thermal-gradient, we focus on the steady-state DW motion speed.
In a Gaussian temperature, the DW cannot immediately follow the local temperature gradient.
Before the DW can reach the steady-state speed corresponding to the local temperature,
it already moves to a position of smaller temperature gradient. More details
about DW motion in Gaussian temperature profile may be an issue of future studies.

\section{Discussion and Summary}

We have studied the thermal gradient-driven DW dynamics in an uniaxial nanowire.
In reality, there is always certain hard anisotropy in a wire whose cross-section
is not a perfect ellipse. Thus, it is interesting to see how the above results
will change in a weak biaxial nanowire with a small hard anisotropy
$K_{\perp}=1/2 \mu_0M^2_s(\mathcal N_z-\mathcal N_y)$, say along $y$-direction.
Our simulations show that a DW still propagates towards the higher temperature
region in a similar way as that in a uniaxial wire. Interestingly, as shown in
Fig. \ref{Fig5.1}(d) for the $K_{\perp}$-dependence of \(v_{\text{simu}}\)
(solid squares) and $d\phi/dt$ (open squares), DW speed increases slightly with
$K_{\perp}$. This may be due to the increase of torque along $\theta$-direction
\cite{AM2007} since $\Gamma_{\theta}$ is proportional to $(\mathcal N_z-\mathcal N_y)$.
This is also consistent with the early results for the uniaxial wire that \(v_{\text
{simu}}\) (which includes stochastic thermal field and demagnetisation fields) is always
larger than \(v_{\text{current}}\) (where the transverse fields are neglected).
At the meanwhile, $d\phi/dt$ decreases with $K_{\perp}$.

The main purpose of this paper is to study the magnonic effects
in thermal-gradient-driven domain wall dynamics. We consider the
spin waves explicitly and all the
material parameters (exchange constant $A$, crystalline anisotropy $K$, 
saturation magnetization $M_s$, and Gilbert damping $\alpha$)
are assumed to be constant. Indeed, the atomistic magnetic moments
are independent of temperature. At the atomistic level, the exchange constant $A$ originating
from the Pauli exclusion principle and the crystalline anisotropy $K$
originating from the spin-orbit coupling only weakly depend on the temperature because of the vibration of atoms
\cite{atomistic}.
In micromagnetic models, because finite volumes that contains many magnetic moments are considered as
unit cells, the parameters $A$, $K$, and $M_s$ depend on the temperature. This is because
the thermally excited spin waves with wavelengths shorter than the length scale of the unit cells 
are included in the effective $A$, $K$, and $M_s$ by doing an average \cite{ABinitio2006,NOWAK11}.
Since we use small mesh size $2\times2\times2$ nm$^3$, only spin waves of very short wavelength 
affect the parameters $A$, $K$, and $M_s$ in our model. Those short-wavelength spin waves possess 
high energy as well as low density of states, so their contributions to the
effective $A$, $K$, and $M_s$ are not significant. The Gilbert damping $\alpha$ depends on the temperature non-monotonically \cite{He1966,Bha1974,Liu2011,Ma2017}. The underlying mechanism is still under debate, 
but for many cases the dependence is not significant in a wide range of temperature. 

In summary, our results show that the uniform thermal gradient always drives a DW
propagating towards the hotter region and the DW-plane rotates around the easy axis. The
DW velocity and DW-plane rotational speed decrease with the damping coefficient.
The DW velocity obtained from simulation agrees with the velocity obtained from
angular momentum conservation when the magnon current density ($J(x)$)
from the simulation is used to estimate the amount of angular momentum transferred
from magnon current to the DW. All the above findings lead to the conclusion that
the thermal gradient interacts with DW through angular-momentum transfer rather than
through energy dissipation. Furthermore, we demonstrated that the magnonic STT
generated by a thermal gradient has both damping-like and  field-like components.
The field-like STT coefficient \(\beta\) is determined from DW speed and DW-plane
rotation speed. \(\beta\) does not depend on the thermal gradient as expected, but
increases with a decrease of DW width. This behavior can be understood from the
expected strong misalignment of magnon spin polarization and the local spin so
that non-adiabatic torque (also called field-like torque) is larger.
For the same reason, a larger Gilbert damping results in a better alignment
between spin current polarization and the local spin, thus \(\beta\) should
decrease with \(\alpha\).  The thermal gradient can be a very
interesting control knob for nano spintronics devices, especially those
made from magnetic insulators.

This work was supported by the National Natural Science Foundation of China
(Grant No. 11774296) as well as Hong Kong RGC Grants Nos. 16300117, 16301518
and 16301816. X.S.W acknowledges support from NSFC (Grant No. 11804045), 
China Postdoctoral Science Foundation (Grant No. 2017M612932 and 2018T110957),
and the Research Council of Norway through its Centres of Excellence funding scheme, Project No. 262633, ``QuSpin.''
M. T. I acknowledges the Hong Kong PhD fellowship.


\newpage
\begin{thebibliography}{99}
	
\bibitem{parkin} Parkin S S P, Hayashi M and Thomas L 2008 \textit{Science}  \textbf{320 190}

\bibitem{DA2005}Allwood D A, Xiong G, Faulkner C C, Atkinson D, Petit D and Cowburn R P 2005 \textit{Science} \textbf{309 1688}
\bibitem{xrw1} Wang X R, P Yan, Lu J and He C 2009 \textit{Ann. Phys. (N. Y.)} \textbf{324 1815}

\bibitem{Fe2009} Wang X R, Yan P and Lu J 2009 \textit{Europhys. Lett.} \textbf{86 67001}

\bibitem{DA2003} Atkinson D, Allwood D A, Xiong G, Cooke M D, Faulkner C C, and Cowburn R P 2003 \textit{Nat. Mater.} \textbf{2 85}

\bibitem{GS2005} Beach G S D, Nistor C, Knutson C, Tsoi M, and Erskine J L 2005 \textit{Nat. Mater.} \textbf{4 741}

\bibitem{MH2006} Hayashi M, Thomas L, Bazaliy Ya B , Rettner C, Moriya R, Jiang X, and  Parkin S S P 2006 \textit{Phys. Rev. Lett.} \textbf{96 197207}

\bibitem{LB1996} Berger L 1996 \textit{Phys. Rev. B} \textbf{54 9353}

\bibitem{Slonczewski} Slonczewski J 1996 \textit{J. Magn. Magn. Mater.} \textbf{159 L1}
\bibitem{SZ2004} Zhang S  and  Li Z 2004 \textit{Phys. Rev. Lett. }\textbf{93 127204}

\bibitem{GT2004} Tatara G  and Kohno H 2004 \textit{Phys. Rev. Lett.} \textbf{92 086601}
\bibitem{Expecom1} Yamaguchi A, Ono T, Nasu S,  Miyake K, Mibu K and  Shinjo T 2004 \textit{Phys. Rev. Lett.} \textbf{92 077205}

\bibitem{Jole2} Hayashi M, Thomas L, Bazaliy Y B, Rettner C, Moriya R, Jiang X, and Parkin S S P  2006 \textit{Phys. Rev. Lett. }\textbf{96 197207}

\bibitem{Expecom2}Yamaguchi A, Hirohata A, Ono T, and Miyajima H  2012  \textit{J. Phys. Condens. Matter} \textbf{24 024201}
\bibitem{XSW11}Yan P, Wang X S, and Wang X R 2011 \textit{Phys. Rev. Lett.} \textbf{107 177207}
\bibitem{NOWAK11} Hinzke D and Nowak U 2011 Phys. Rev. Lett. \textbf{107 027205}
\bibitem{KOVALEV12}Kovalev A A  and Tserkovnyak Y 2012 \textit{Europhys. Lett.} \textbf{97 67002}
\bibitem{MagSTT}Wang X R, Yan P and Wang X S 2012 IEEE Trans. Magn. \textbf{48 11}.
\bibitem{spincal}Bauer G E W, Saitoh E and van Wees B J  2012 \textit{Nat. Mater}. \textbf{11 391}
\bibitem{Jiang13} Jiang W, Upadhyaya P, Fan Y B, Zhao J, Wang M S, Chang L T, Lang M R, Wong K L, Lewis M, Lin Y T, Tang J S, Cherepov S,  Zhou X Z, Tserkovnyak Y, Schwartz R N and Wang K L 2013 \textit{Phys. Rev. Lett.} \textbf{110 177202}

\bibitem{NOWAK14}Schlickeiser F, Ritzmann U, Hinzke D and Nowak U 2014 \textit{Phys. Rev. Lett.} \textbf{113 097201}

\bibitem{XSW14} Wang X S  and Wang X R 2014 \textit{Phys. Rev. B} \textbf{90 014414}
\bibitem{XGWANG16} Wang X-G, Chotorlishvili L, Guo G-H, Sukhov A, Dugaev V, Barnas J and Berakdar J  2016
              Phys. Rev. B \textbf{94 104410}

\bibitem{Nat.com2017} Safranski C, Barsukov I, Lee H K, Schneider T, Jara A, Smith A, Chang H, Lenz K, Lindner J,Tserkovnyak Y, Wu M and Krivorotov I  2017 \textit{Nat. Com.} \textbf{8 117}



\bibitem{XGWANG12}Wang X G, Guo G H, Nie Y Z, Zhang G F and Li Z X 2012 \textit{Phys. Rev. B} \textbf{86 054445}

\bibitem{PYAN2015}Yan P, Cao Y, and Sinova J 2015 \textit{Phys. Rev. B} \textbf{92 100408}


\bibitem{Nonad}	Kishine Jun-ichiro  and Ovchinnikov A S 2010 \textit{Phys. Rev. B} \textbf{81 134405}



\bibitem{Brown1963}Brown W F 1963 \textit{Phys. Rev. } \textbf{130 1677}

\bibitem{TLGIL}Gilbert T L 2004 \textit{IEEE Trans. Magn.} \textbf{40 3443}


\bibitem{Mumax}Vansteenkiste A, Leliaert J, Dvornik M, Helsen M, Garcia-Sanchez F
and Van Waeyenberge B 2014 \textit{AIP Advances} \textbf{4 107133}

\bibitem{Nowak2008} Hinzke D, Kazantseva N, Nowak U, Mryasov O N, Asselin P and Chantrell R W 2008
\textit{Phys. Rev. B} \textbf{77 094407}

\bibitem{spintensor} Etesami S R, Chotorlishvili L, Sukhov A and Berakdar J  2014
\textit{Phys. Rev. B} \textbf{90 014410}

\bibitem{AM2007} Mougin A, Cormier M, Adam J P, Metaxas P J  and  Ferre J 2007 \textit{Europhys. Lett.} \textbf{78 57007}

\bibitem{RW2010}Wieser R, Vedmedenko E Y, Weinberger P and Wiesendanger R 2010 \textit{Phys. Rev. B} \textbf{82 144430}

\bibitem{Euro}Sun Z Z, Schliemann J, Yan P and Wang X R  2011 \textit{Eur. Phys. J. B} \textbf{79 449–453}

\bibitem{exp2015}Ramsay A J, Roy P E, Haigh J A, Otxoa R M, Irvine A C, Janda T, Campion R P, Gallagher B L and Wunderlich J 2015
\textit{Phys. Rev. Lett}. \textbf{114 067202}

\bibitem{Gaussian2017}Moretti S, Raposo V, Martinez E and Lopez-Diaz L 2017
\textit{Phys. Rev. B} \textbf{95 064419}

\bibitem{ABinitio2006}Staunton J B, Szunyogh L, Buruzs A, Gyorffy B L, Ostanin S, and L. Udvardi
2006
\textit{Phys. Rev. B} \textbf{74 144411}

\bibitem{atomistic}Chico J, Etz C, Bergqvist L, Eriksson O, Fransson J, Delin A, and Bergman A
2014 \textit{Phys. Rev. B} \textbf{90 014434}


\bibitem{He1966}Heinrich B,   Frait Z 1966 \textit{Phys. Status Solidi B} \textbf{16 K11}
\bibitem{Bha1974} Bhagat S M,  and Lubitz P 1974 \textit{Phys. Rev. B} \textbf{10 179}
\bibitem{Liu2011}  Liu Y, Starikov A A,  Yuan Z, and Kelly P J 2011  \textit{Phys. Rev. B} \textbf{84 014412 }
\bibitem{Ma2017} Maier-Flaig H, Klingler S , Dubs C, Surzhenko O, Gross R, Weiler M,  Huebl H, and Goennenwein S T B 2017 \textit{Phys. Rev. B} \textbf{95 214423}

\end{thebibliography}
\end{document}